\documentclass[aps,prl,twocolumn,showpacs,amsmath,amssymb]{revtex4}

\usepackage{graphicx}
\usepackage{dcolumn}
\usepackage{bm}

\begin{document}

\title{Microscopic structure of liquid hydrogen: a neutron diffraction experiment}
\author{M. Celli$^1$, U. Bafile$^1$, G.J. Cuello$^2$, F. Formisano$^3$, E. Guarini$^4$, R. Magli$^{4,5}$, M. Neumann$^6$, and M. Zoppi$^1$}
\affiliation{
$^1$ Consiglio Nazionale delle Ricerche, Istituto di Fisica Applicata "Nello Carrara",\\
Via Panciatichi 56/30, I-50127 Firenze, Italy.\\
$^2$ Institut Laue-Langevin, 6 rue J. Horowitz, BP 156, F-38042, Grenoble Cedex 9, France.\\
$^3$ Istituto Nazionale per la Fisica della Materia - Operative Group in Grenoble,\\
c/o Institut Laue-Langevin, 6 rue J. Horowitz, F-38042, Grenoble Cedex 9, France.\\
$^4$ Istituto Nazionale per la Fisica della Materia, Unit\`a di Firenze,\\
Via G. Sansone 1, I-50019 Sesto Fiorentino, Italy.\\
$^5$ Dipartimento di Chimica e Biochimica Medica, Universit\`a di Milano,\\
Via F.lli Cervi 93, I-20090 Segrate, Italy.\\
$^6$ Institut f\"ur Experimentalphysik der Universit\"at Wien, Strudlhofgasse 4, A-1090 Wien, Austria.\\
}
\date{\today}

\begin{abstract}
We have measured the center-of-mass structure factor $S(k)$ of liquid para-hydrogen by neutron diffraction, using the D4C diffractometer at the Institute Laue Langevin, Grenoble, France.
The present determination is at variance with previous results obtained from inelastic neutron scattering data, but agrees with path integral Monte Carlo simulations.

\end{abstract}

\pacs{67.20.+k, 61.20.Ja, 61.12.-q}

\maketitle

Among the simple liquids, the microscopic structure factor, $S(k)$, of liquid hydrogen has been one of the most jealously kept secrets of nature.
This is a consequence of several experimental problems.
Due to the small number of electrons, an X-ray determination of the hydrogen structure factor is not an easy task.
On the other hand, a neutron diffraction experiment is particularly difficult because of the relevance of inelastic scattering events: their size, mainly determined by the ratio between the neutron and the nuclear mass \cite{Placzek}, makes the standard correction techniques unsuitable to hydrogen.
Due to the doubled nuclear mass, the experiment analysis is less demanding for deuterium.
Nonetheless, the first reliable neutron diffraction measurement of the microscopic structure of liquid deuterium is relatively recent and was carried out using a small-angle, time-of-flight, diffractometer (SANDALS, at ISIS, UK) that was specifically built for light mass liquids.
Using this instrument, the structure factor of liquid $\rm{D_2}$ was measured in the vicinity of the triple point \cite{D2_tp} and close to the freezing transition \cite{D2_ml}.
By combining the data of Ref.~\onlinecite{D2_tp} with those measured in a further diffraction experiment, performed on a reactor source (7C2 diffractometer at Laboratoire L\'eon Brillouin, Saclay, France) \cite{D2_7c2}, an improved determination of the structure factor of liquid deuterium was obtained.
This turned out to be in excellent agreement with path integral Monte Carlo (PIMC) simulation results \cite{zoppi95}.

The knowledge of $S(k)$ for deuterium is not easily transferred to hydrogen, due to the different role played by quantum effects \cite{DeBoer}.
On the other hand, for the case of hydrogen, the overwhelming ratio between the incoherent and coherent neutron scattering cross section of the proton makes it extremely difficult to extract the intermolecular response, which carries the structural information, from the large intramolecular contribution.
The first successful attempt to obtain structural information on liquid hydrogen, carried out on SANDALS, allowed us to determine the thermodynamic derivatives of $S(k)$ \cite{ZCS98}.
These were found in a rather good quantitative agreement with the PIMC simulation results \cite{lviv01}.

Due to the well recognized experimental difficulties, alternative methods have been suggested to determine $S(k)$ for liquid hydrogen.
Bermejo and co-workers \cite{bermejo} have reported a determination of the structure factor of liquid para-hydrogen using the results of an inelastic neutron scattering experiment and the sum rule which relates the dynamic structure factor, $S(k,\omega)$, to the static one, $S(k)$, \cite{SWL}:

\begin{equation}
\label{S(Q)}
S(k) = \int ^{+\infty} _{-\infty} d{\omega} \: S(k,\omega).
\end{equation}
The results were qualitatively reasonable, but not fully convincing on a quantitative basis \cite{zoppi_02}.
In fact, the main peak of $S(k)$ appeared rather high, exceeding the value of 2.85 which marks, according to the Hansen--Verlet criterion \cite{hansen_verlet}, the onset of the freezing transition.
Even though Hansen and Verlet formulated this rule for simple model systems (classical monatomic particles interacting through a Lennard-Jones potential) it is hard to believe that for hydrogen, a genuine quantum system for which one expects an overall broadening and damping of the structural features, $S(k)$ should reach such a large value in the liquid phase.

More recently, the same method, i.e. integrating the dynamic structure factor at constant $k$, was applied by Pratesi {\it et al.} \cite{pratesi} using the X-ray inelastic scattering data obtained on the ID16 beamline at the European Synchroton Radiation Facility (Grenoble, France).
Their results for $S(k)$, taken at $T=31.5$ K and molecular number density $n=$21.5~nm$^{-3}$, were not much extended in $k$, barely exceeding the main peak position of the structure factor, and with rather large error bars.
However, the peak height turned out considerably lower than the neutron data of Ref. \onlinecite{bermejo} and consistent with their PIMC simulation results.
The X-ray diffraction measurement, carried out in parallel \cite{pratesi}, showed much smaller error bars and still a good agreement with the PIMC simulation.
However, in this case, the accessed $k$-range did not even reach the peak position of $S(k)$.

In this context, we have carried out a new experiment aiming to determine directly, for the first time by neutron diffraction, the center-of-mass structure factor of liquid para-hydrogen and the results constitute the object of the present letter.
The measurements were carried out using the D4C diffractometer of the Institut Laue Langevin (Grenoble, France) in its standard configuration with an incident wavelength $\lambda_0 =0.6933$ \AA.
Of the four thermodynamic states investigated, we report here the results at $T=17.1\pm 0.1$ K and $p=29.9\pm 0.1$ bar ($n=22.95 \pm 0.05$ nm$^{-3}$ \cite{pvt}).
The liquid hydrogen sample was condensed directly in a vanadium cylindrical container (6 mm internal diameter, 0.2 mm wall thickness) imbedded in the low pressure buffer gas of an {\it orange} cryostat.
At the bottom of the vanadium container, out of the neutron beam, we had inserted some powder of a paramagnetic catalyst made of Cr$_2$O$_3$ on an Al$_2$O$_3$ substrate, in order to speed up the conversion from ortho- to para-hydrogen.
The relative concentration of the two species was monitored looking at the low momentum transfer portion of the diffraction pattern \cite{ZCS98}.
Twenty-four hours of conversion time were allowed to stabilize the sample.
Then, the stability of the low-$k$ portion of the diffraction pattern confirmed that the sample had reached the thermodynamic equilibrium concentration, calculated to be 99.96\% rich in para-hydrogen.
In order to check the overall stability of the experimental setup, as well as of the sample, for each thermodynamic point we took several measurements in repeated runs.

\begin{figure}
\resizebox{0.5\textwidth}{!}{%
\includegraphics[angle=0]{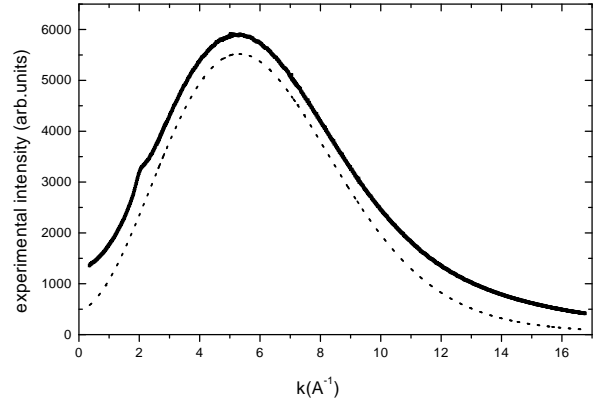}
}
\caption{Diffraction pattern of liquid para-hydrogen at $T=17.1$ K and $p=29.9$ bar.
What in the figure appears as a bold line is composed, instead, by the ensemble of the experimental points.
The lack of intensity in the low-$k$ region is peculiar of the para species.
The decay at high momentum transfer is determined by the finite energy of the incident neutrons (see text).
The tiny structure in the region $k \simeq 2$ \AA$^{-1}$ is the signature of the main peak of the intermolecular structure factor $S(k)$.
The dashed line is the calculated self molecular part.}
\label{fig_1}
\end{figure}

In order to properly subtract the background contribution from the raw data, we have carried out an additional measurement, without changing the experimental setup and filling the sample container with a small amount of gaseous $^3$He.
The density of the gas was chosen such as to match the scattering power of the liquid para-hydrogen sample with the absorption power of $^3$He.
The details of the procedure can be found in Ref.~\onlinecite{calib3He}.
We used the value $\sigma(\lambda_0)=43.4$ barn, obtained from Ref.~\onlinecite{czrs99}, for the molecular hydrogen scattering cross section at the incident neutron wavelength.
In Fig.~1, we show the diffraction pattern of liquid para-hydrogen after subtraction of background and container scattering, properly corrected for attenuation.
The diffraction spectrum is dominated by the {\it self} molecular part, which is produced by the intramolecular structure.
The {\it distinct} coherent contribution, containing the required information on the intermolecular structure factor, appears as a tiny undulation superimposed on the much larger intramolecular structure in the region $k \simeq 2$ \AA$^{-1}$.
The difference between the experimental and calculated patterns (see Fig.~1) is mainly due to multiple scattering effects.

The data treatment needed to extract the structure factor requires the calculation of the self part of the molecular double differential neutron cross section of para-hydrogen.
This can be done using the Young and Koppel (YK) theory \cite{YK}.
Here, the intermolecular interactions are totally neglected and the hydrogen molecules are modeled as a set of non-interacting particles possessing their relevant internal degrees of freedom.
Thus, each molecule is considered separately and the intramolecular roto-vibrational modes, and spin correlations, are explicitly taken into account.
The vibrational modes are considered harmonic and the rotations are free.
This model applies well to our case because, at low temperatures, only the ground vibrational state is populated and, using thermal neutrons, no vibrational transition is allowed.
Moreover, as long as the system is in the liquid phase and is not subject to high pressures, the anisotropic components of the intermolecular potential are negligible \cite{VanK,Z_96,MSUZ}.
The vibrational-rotational coupling can be accounted for, in an effective way, using the experimental values for the roto-vibrational levels \cite{H&H}.
Within the applicability limits of this model, a rigorous calculation of the double differential neutron cross section is possible.

It is well known \cite{Ceperley95} that for hydrogen, due to quantum effects, the average center-of-mass kinetic energy is different from the classical value $\langle E_{\rm k} \rangle = (3/2)k_{\rm B}T$ and depends on density \cite{langel88}.
This quantity can be directly measured from an inelastic neutron scattering experiment \cite{CCZ00}.
We have shown \cite{czrs99} that a modified Young and Koppel (MYK) model can be defined to account for this property, using an effective temperature in the width of the ideal-gas Gaussian that describes the dynamic structure factor of the molecular centers of mass.
By means of the MYK model, the {\it self} portion of the double differential neutron cross section could be integrated, at each scattering angle $\theta$, over the energies of the scattered neutrons.
The result, properly normalized as described below, is represented by the dashed line in Fig.~1.

As far as the multiple scattering effects are concerned, we carried out a Monte Carlo (MC) simulation using, as input, the calculated MYK cross section.
We found that the multiple scattering contribution to the cross section is a rather smooth function of $k$.
However, since its intensity is larger than the intermolecular part (see Fig.~1), the accuracy of a typical multiple scattering calculation might be insufficient for a reliable extraction of $S(k)$.
On the other hand, the similarity between the measured diffraction pattern and the calculated intramolecular cross section suggests to consider an overall, rather unstructured, generalized background, that includes the multiple scattering and should be represented by a low-order polynomial function $P(k)$.
Therefore, instead of attempting a difficult evaluation of the multiple scattering, we applied a different procedure where the order and coefficients of the polynomial, as well as the overall normalization constant, were obtained from a fit.
A model function, including $P(k)$ and the calculated intramolecular diffraction pattern, was fitted to those data points measured at $k-$values where an independent knowledge of the intermolecular structure factor is available: namely, we used $S(k) \simeq S(0)$ at low $k$, as is typical of a dense liquid (with $S(0)$ related to the isothermal compressibility), and $S(k) \simeq 1$ for large $k$.
The $k$-range that contains the experimental information on the sought-for structure factor was excluded from the fit.
We found that a fifth order polynomial is sufficient to give a satisfactory description of the generalized background function which turned out to be in a semiquantitative agreement with the MC evaluation of the multiple scattering intensity.

\begin{figure}
\resizebox{0.5\textwidth}{!}{%
\includegraphics[angle=0]{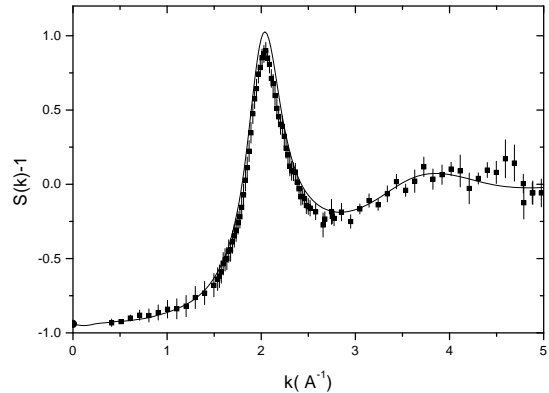}
}
\caption{Measured intermolecular (i.e. center-of-mass) structure factor of liquid para-hydrogen at $T=17.1$ K and $p=29.9$ bar.
The error bars are determined as described in the text.
Data are rebinned every five points in the region where $S(k)$ is weakly $k-$dependent (i.e. for $k<1.5$ \AA$^{-1}$ and $k>2.8$ \AA$^{-1}$).
The full line represents the results of a PIMC simulation using the Norman {\it et al.} \protect\cite{NWB} spherical potential.
The value for $S(k=0)$ is obtained from the isothermal compressibility \protect\cite{pvt}.
}
\label{fig_2}
\end{figure}

The fitted polynomial has been subtracted, together with the calculated intramolecular contribution, from the experimental diffraction pattern.
The result gives, after normalization, the intermolecular term in the measured cross section that can be written as:

\begin{equation}
\left ( \frac{d\sigma}{d\Omega} \right )_{\rm{coh}}=A_{s,sc}(k) u(k) [S(k)-1],
\end{equation}
where $A_{s,sc}$ is the Paalman and Pings \cite{P&P} attenuation coefficient and $u(k)$ is the molecular form factor \cite{zoppi93}.
It is worthwhile to note that, in the data analysis, we have taken into account self absorption as well as the finite dimensions of sample and detectors.
The final result for $S(k)$ is reported in Fig.~2.
The error bars account for both the propagated statistical errors from the raw spectrum and an estimate of the systematic errors coming from the fitting procedure.
For example, we have taken into account, among other things, the effect of the variation of the fitting intervals in the determination of the polynomial background.
In the same figure, we compare the present experimental neutron diffraction results with the corresponding quantity obtained from a PIMC computer simulation \cite{zoppi_02} using the Norman, Watts, and Buck \cite{NWB} (NWB) intermolecular potential.
The simulation conditions were $T=17.1$ K and $n=23.0$ nm$^{-3}$.
The agreement is not perfect but still is rather good and both the experimental and simulation data consistently show no trace of an anomalously high main peak in $S(k)$, as it might have been inferred from Ref.~\onlinecite{bermejo}.

We compare the present results with a corresponding analysis carried out on liquid deuterium, in similar thermodynamic conditions and using the same intermolecular potential (NWB).
In that case, the agreement between simulation and experiment was extremely good, and data were of such a quality to allow a good discrimination among different intermolecular interaction functions, namely NWB and Lennard-Jones \cite{zoppi95}.
In the present case, an overall quantitative agreement with simulation is also found, although the experiment is much more difficult.
Consequently, the resulting uncertainties are larger, and do not allow us to clearly discriminate among various choices of the intermolecular potential.

In order to be more specific, we examined the structural information that is obtained from the simulations, using two different potential functions.
To this aim, we compared the present simulation data at $T=17.1$ K and $n=22.22$ nm$^{-3}$, based on the NWB potential, with existing PIMC data at $T=17.2$ K and $n=22.1$ nm$^{-3}$ \cite{CCZ}, obtained using the Silvera and Goldman (SG) intermolecular potential \cite{S&G}.
The only observed difference was a lower main peak, by $\simeq 4\%$ in $[S(k)-1]$, for the simulation results based on the SG potential function.
Considering that temperature and density differ, in the two simulations, by $\simeq 0.6\%$ and $\simeq 0.5\%$ respectively, we can mainly attribute to the different choice of the interaction potential the found deviation of $\simeq 4\%$.
This, in turn, is a little smaller than the present size of the error bars in the main peak region.

Finally, on closer inspection of Fig.~2, data seem to suggest a slightly narrower main peak than in the simulation.
We exclude that such an effect could be assigned to the particular choice of the intermolecular potential.
In fact, by comparing the NWB and the SG simulations, no observable difference in the main-peak width could be evidenced.
Of course, we could attribute this effect to some remaining systematic experimental error that we were unable to get rid of.
However, it is interesting to note that a similar qualitative result, concerning the width of $S(k)$, was also found in Ref.~\onlinecite {bermejo} and attributed to the presence, in the real liquid, of significantly longer range correlations than those depicted by the simulation.

\section{Acknowledgments}
\begin{acknowledgments}
The skillful technical assistance of Mr. Pierre Palleau (ILL, Grenoble, France), during the setup of the experiment, is gratefully acknowledged.
\end{acknowledgments}


\begin{thebibliography}{}

\bibitem{Placzek} G. Placzek, Phys. Rev. {\bf 86}, 377 (1952).
\bibitem{D2_tp} M. Zoppi, U. Bafile, R. Magli, and A. K. Soper, Phys. Rev. E {\bf 48}, 1000 (1993).
\bibitem{D2_ml} M. Zoppi, A. K. Soper, R. Magli, F. Barocchi, U. Bafile, and N. W. Ashcroft, Phys. Rev. E {\bf 54}, 2773 (1996).
\bibitem{D2_7c2} E. Guarini, F. Barocchi, R. Magli, U. Bafile, and M. C. Bellissent-Funel, J. Phys.: Condens. Matter {\bf 7}, 5777 (1995).
\bibitem{zoppi95} M. Zoppi, U. Bafile, E. Guarini, F. Barocchi, R. Magli, and M. Neumann, Phys. Rev. Lett. {\bf 75}, 1779, (1995).
\bibitem{DeBoer} J. de Boer, Rep.\ Prog.\ Phys.\ {\bf 12}, 305 (1949).
\bibitem{ZCS98} M. Zoppi, M. Celli, and A. K. Soper, Phys. Rev. B {\bf 58}, 11905 (1998).
\bibitem{lviv01} M. Zoppi, M. Celli, U. Bafile, E. Guarini, and M. Neumann, Condens. Matter Phys. {\bf 4}, 283 (2001).
\bibitem{bermejo} F. J. Bermejo, K. Kinugawa, C. Cabrillo, S. M. Bennington, B. F\protect{\aa}k, M. T. Fern\'andez-D\'iaz, P. Verkerk, J. Dawidowski, and R. Fern\'andez-Perea, Phys. Rev. Lett. {\bf 84}, 5359 (2000).
\bibitem{SWL} S. W. Lovesey, {\it Theory of Neutron Scattering from Condensed Matter}, Vol.~1 (Oxford University Press, Oxford, 1987).
\bibitem{zoppi_02} M. Zoppi, M. Neumann, and M. Celli, Phys. Rev. B {\bf 65}, 092204 (2002).
\bibitem{hansen_verlet} J. P. Hansen and L. Verlet, Phys. Rev. {\bf 184}, 151 (1969).
\bibitem{pratesi} G. Pratesi, D. Colognesi, A. Cunsolo, R. Verbeni, M. Nardone, G. Ruocco, and F. Sette, Phil. Mag. B {\bf 82}, 305 (2002).
\bibitem{pvt} H. M. Roder, G. E. Childs, R. D. McCarty, and P. E. Angerhofer, NBS Technical Note n. 641 (1973).
\bibitem{calib3He} F. Barocchi, P. Chieux, R. Fontana, R. Magli, A. Meroni, A. Parola, L. Reatto, and M. Tau, J. Phys.: Condens. Matter {\bf 9}, 8849 (1997).
\bibitem{czrs99} M. Celli, M. Zoppi, N. Rhodes, and A. K. Soper, J. Phys.: Condens. Matter {\bf 11}, 10229 (1999).
\bibitem{YK} J. A. Young and J. U. Koppel, Phys. Rev. A {\bf 135}, 603 (1964).
\bibitem{VanK} J. Van Kranendonk, {\it Solid Hydrogen} (Plenum, New York, 1983).
\bibitem{Z_96} M. Zoppi, L. Ulivi, M. Santoro, M. Moraldi, and F. Barocchi, Phys. Rev. A {\bf 53}, R1935 (1996).
\bibitem{MSUZ} M. Moraldi, M. Santoro, L. Ulivi, and M. Zoppi, Phys. Rev. B {\bf 58}, 234 (1998).
\bibitem{H&H} K. P. Huber and G. Herzberg, {\it Constants of Diatomic Molecules} (Van Nostrand, New York, 1979).
\bibitem{Ceperley95} D. M. Ceperley, Rev.\ Mod.\ Phys.\ {\bf 67}, 279 (1995).
\bibitem{langel88} W. Langel, D. L. Price, R. O. Simmons, and P. E. Sokol, Phys. Rev. B {\bf 38}, 11275 (1988).
\bibitem{CCZ00} M. Celli, D. Colognesi, and M. Zoppi, Eur. Phys. J. B {\bf 14}, 239 (2000).
\bibitem{P&P} H. H. Paalman and C. J. Pings, J. Appl. Phys. {\bf 33}, 2635 (1962).
\bibitem{zoppi93}M. Zoppi, Physica B {\bf 183}, 235 (1993).
\bibitem{NWB} M.J. Norman, R.O. Watts, and U. Buck, J. Chem. Phys. {\bf 81}, 3500 (1984).
\bibitem{CCZ} M. Celli, D. Colognesi and M. Zoppi, To be published.
\bibitem{S&G} I. F. Silvera, V. V. Goldman, J. Chem. Phys. {\bf 69}, 4209 (1978).

\end{thebibliography}
\end{document}